\documentclass[12pt]{article}
\usepackage{amssymb}

\begin{document}

\begin{center}
\bf INFORMATION TRANSFER IN QUANTUM MEASUREMENTS:
\bf IRREVERSIBILITY AND AMPLIFICATION  \footnote{Lecture Notes from 
NATO Advanced Study Institute held in Bad Windsheim, West Germany, 
August 17-30, 1981.  Published in the Proceedings,
{\it Quantum Optics, Experimental Gravitation, and Measurement
Theory}, edited by P. Meystre and M. O. Scully (Plenum Press, 1983), pp.
87-116.  The
text is unaltered, except for minor changes necessary to input it in Tex.}
\end{center}

\begin{center} Wojciech Hubert Zurek \footnote{Present address:  Theory
Division, T-6, MS B288, Los Alamos National Laboratory, P.O. Box 1663, Los
Alamos, NM 87545}\\ \vspace*{0.15in} Theoretical Astrophysics 130-33 \\
California Institute of Technology \\ Pasadena, CA 91125 \end{center}

{\it{[Heisenberg] ... remarks ... that even in the case of macroscopic
phenomena we may say, in a certain sense, that they are created by repeated
observations ... }}

[Niels Bohr$^1$, in 1928 \underline{Nature} article]

\noindent
{\bf I. \hspace{.15 in} INTRODUCTION}

  The aim of these lectures is to investigate the transfer of information
occurring in course of quantum interactions.  In particular, I shall
explore circumstances in which such an information transfer with the
quantum environment of the considered quantum system leads to the
destruction of the phase coherence between the states of the privileged
basis in the system Hilbert space.  This basis shall be called the
\underline{pointer basis}.  I shall argue that states of this pointer
basis correspond to the ``classical'' states of the observables of the
quantum system in question. 

The pointer basis emerges as a consequence of the environment-induced
superselection rules:  the system cannot be observed in a superposition of
the states of the pointer basis, because the environment continuously
performs a non-demolition measurement of the observable diagonal in the
pointer basis.  Applied to the quantum apparatus, the environment-induced
superselection rules single out a preferred basis of the apparatus
pointer.  The apparatus can be observed only in one of the eigenstates of
the pointer basis. This fact provides a natural explanation of the apparent
reduction of the state vector.  The increase of entropy associated with the
process of reduction is then a consequence of the transfer of information
into the environment degrees of freedom.  Amplification of the measurement
outcome facilitates this effectively irreversible information transfer and
increases reliability of the correlation between the states of the pointer
basis and the state of the measured system.

\noindent
{\bf II. \hspace{.15 in} TWO STAGES OF THE MEASURENENT}

The state vector of the measured quantum system becomes, as a result of the
ideal measurement, one of the eigenstates of the measured operator.  The
choice of the particular, single eigenstate $| i \rangle $ and of the
corresponding
eigenvalue $\lambda_i$  appears to be fundamentally stochastic.  It is
governed
by the thoroughly tested probabilistic law

\{probability of $|s \rangle $ to become $|i \rangle \} = p (|s \rangle
\rightarrow |i \rangle ) = | \langle s |i \rangle |^2$
.

\noindent
Yet the evolution equation for the state vector, e.g. the Schr\"{o}dinger
equation in nonrelativistic quantum mechanics, is linear and its solutions
are deterministic and satisfy the principle of superposition.  As a result
the inherently random process of measurement culminating in the ``collapse
of the wave function'' or ``reduction of the state vector'' cannot be fitted
within the dynamical evolution of a \underline{closed} quantum system.  This
discord
between the apparently stochastic course of measurements on the one hand
and the deterministic evolution of the state vectors on the other has been
in the center of attention since the early days of the modern quantum
theory. This is hardly surprising, as the process of measurement is of
great importance by itself.  Moreover, the random nature of the collapse
allows one to expect that understanding it will shed a new light on the
long-standing problems of statistical mechanics concerning irreversibility
and the concept of entropy. This in turn should explain how the
counterintuitive quantum phenomena can conspire to create the everyday
classical, macroscopic reality.
\newpage

The goal of this paper is to explore the idea that the measuring apparatus
as well as other classical, macroscopic systems violate the quantum
superposition principle because under normal conditions their observables
interact with \underline{the external universe}.  Therefore, macroscopic
systems
cannot
be regarded as ``closed''.  To argue this I shall show how the effective
superselection rules can be induced by the environment:  ``Monitoring'' of
the state of the system by the environment will be shown to result in the
apparent collapse of the wave packet.  Much of the discussion given below
is an elaboration of the recently introduced idea of the pointer basis.  It
should be noted that the idea of macroscopic observables arising somehow in an
open system observed by the environment has been recently expressed by Zeh$^8$
as well as by Wigner$^9$.\footnote{See, in particular, p. 268 of Ref. 8 and
Eq.
(27) of Ref. 9.}

I shall begin the discussion of the measurement process by reviewing von
Neumann's approach$^3$, who, in his analysis of the measurement, distinguishes
two fundamentally different stages:

\noindent
\underline{Establishment of the Correlation}

During this first stage the state vector of the quantum apparatus, initially
$|A_o\rangle$, becomes correlated with the state vector $|\psi\rangle$ of the
measured
system
$$
	|\Phi_i\rangle  =  |A_o\rangle \otimes | \psi\rangle = \left( \sum_{n}
	\alpha_n|A_n\rangle \right)
	\otimes \left(\sum_{s} c_s|s\rangle \right) \longrightarrow $$

	$$
	\longrightarrow  \sum_{s} c_s |A_s\rangle \otimes | s\rangle = |
\Phi_f\rangle  .
	 \eqno(2.1)
$$

Here $\{|A_s\rangle\}$ and $\{|s\rangle\}$ constitute orthonormal bases in the
apparatus and
system Hilbert spaces, respectively.  Such correlation can be accomplished
by an
appropriate interaction hamiltonian $H^{AS}$ acting over a time interval
$\tau$:
$$
|\Phi_f\rangle = exp(-iH^{AS} \tau / \hbar )|\Phi_i\rangle = U|\Phi_i\rangle .
\eqno(2.2)
$$
If the system in the eigenstate of some operator $\hat{S} =
\sum_s\lambda_s|s\rangle\langle s|$  is not to be perturbed by the measurement,
$H^{AS}$ with
$\hat{S}$:
$$
\left[H^{AS}, \hat{S}\right] = 0 \ \ .
\eqno(2.3)
$$
Above commmutation relation must be always satisfied by
such``nonperturbative'' or ``nondemolition'' measurements,$^{11-15}$ which
will
be of crucial importance for our further discussion.

The most important feature of a successful measurement---a one-to-one
correlation between the state of the apparatus and the state of the
measured system---is already present in the wave-function given by Eq.
(2.1).  There is, however, a fundamental difficulty caused by the fact that
the final state given by that equation is pure, and not the appropriate
mixture.  This difficulty is best appreciated when we construct the density
matrix corresponding to the final pure state $|\Phi_f\rangle$;
$$
\rho_{pure} = |\Phi_f\rangle \langle\Phi_f| = \sum_{s} \sum_{s\prime}
c^*_{s\prime}
c_s|A_s\rangle \langle A_{s\prime} |\otimes| s\rangle \langle s\prime| \ \ ,
\eqno(2.4)
$$
and compare it with the mixture diagonal in $|A_s\rangle \langle A_s |
\otimes|
s\rangle \langle s |$:
$$
\rho_{mix} = \sum_{s} |c_s|^2 |A_s\rangle \langle A_s | \otimes | s\rangle
\langle s| \eqno(2.5)
$$
It is this $\rho_{mix}$ which contains probabilities of different outcomes,
$|A_s\rangle \langle A_s|\otimes| s\rangle \langle s|$, with the
appropriate weights.  The density
matrix $\rho_{pure}$ contains, apart from the diagonal terms which are
identical with $\rho_{mix}$, also off-diagonal ``correlations''.  Their
presence results in different physical properties of the ensembles
described by $\rho_{pure}$ and $\rho_{pure}$ In particular:

(i) \hspace{.15 in} The state described by $\rho_{pure}$ can be brought
back to the initial state, $ |A_o\rangle \otimes | \psi\rangle$, by means of
hamiltonian evolution.

(ii) \hspace{.15 in} The apparatus contains not only complete information
about the measured observable $\hat{S}$, but also information about other
observables which do not commute with $\hat{S}$.  This can be exhibited by
rewriting $|\Phi_f\rangle$ in terms of an alternative orthonormal apparatus
basis:
$$
|\Phi_f\rangle = \sum_{s} c_s| A_s\rangle \otimes | s\rangle = \sum_{p} b_p |
A_p\rangle \otimes
| p\rangle \ \ . \eqno(2.6)
$$
Here $|A_p\rangle = \sum_s \langle A_s | A_p\rangle |A_s\rangle $.  The
relative
states$^{16} \ |p\rangle
= (b_p)^{-1} \sum_s \langle A_p | A_s\rangle | s\rangle $ are normalized, but,
in the
generic case, not necessarily orthogonal. They can nevertheless be used as
``eigenstates'' to define an observable which is non-hermitean, but has
many of the properties of hermitean operators.$^{17}$  Therefore, it is not
clear what is the state vector in which the system has been registered by
the apparatus.  The ensuing problems are particularly apparent when all the
coefficients $c_s$ in Eq. (2.1) happen to be equal, and thus the set
$\{|p\rangle\}$ of the relative basis states is orthonormal. Then the
apparatus
correlated with the measured system contains all the information not only
about the observable $\hat{S}$, but also about the observable $\hat{P} =
\sum_p \gamma_p | p\rangle \langle p|$.  This is so despite the fact that
$\hat{S}$ and
$\hat{P}$ do not  commute!  Yet we know that quantum mechanics prevents one
from simultaneously measuring two noncommuting observables.  The quantum
apparatus seems to ignore this law, following directly from the principle
of indeterminacy!  Furthermore, everyday experience convinces us that the
choice of what the apparatus has measured cannot be made by us, arbitrarily
long after the apparatus-system interaction has ceased, as the reasoning
above seems to imply.  To eliminate such difficulties von Neumann has
postulated the occurrence of a second stage of the measurement process:

\noindent
\underline{Reduction of the State Vector}

During this second stage of the measurement process the density
matrix  $\rho_{pure}$ of the combined apparatus-system object undergoes the
transition:
$$
	\rho_{pure} =  \sum_{s} \sum_{s\prime} c_{s\prime}^* c_s | A_s\rangle
\langle A_{s\prime}
	|\otimes| s\rangle \langle s\prime|  \longrightarrow $$

	$$
	\longrightarrow  \sum_{s} |c_s|^2 |A_s\rangle \langle A_s |\otimes |
s\rangle\langle s| =
\rho_{mix} \ \  .
	 \eqno(2.7)
$$
It is usually assumed that both the states of the apparatus and the states
of the system are mutually orthonormal.  However, as we have already
discussed, states of the system relative to the privileged pointer basis
states of the apparatus are, in general, not necessarily orthogonal.$^{17}$

Reduction of the state vector cannot be accomplished in a closed system by
means of unitary evolution.  This is a straight forward consequence of the
fact that unitary transformations evolve state vectors into state
vectors.  Hence, they are inherently incapable of transforming
$\rho_{pure}$, a projection operator:
$$
\left(\rho_{pure}\right)^2 = \rho_{pure}  \eqno(2.8)
$$
into $\rho_{mix}$,  for which:
$$
Tr\left(\rho_{mix} \right)^2 < Tr \rho_{mix} \eqno(2.9)
$$
To accomplish such transformation, one must employ a four-index object
called ``superoperator''. Needless to say, objects like that cannot exist in
the dynamics governed by Schr\"{o}dinger equation. (Their existence has
been nevertheless considered both in the context of statistical
mechanics$^{18}$ and, more recently, in the field of black hole
thermodynamics.$^{19, 20}$)

Von Neumann was of course well aware of this dilemma. He favored the view
according to which the conciousness of the observer was the ultimate cause
of the collapse.  This view was subsequently elaborated by London and
Bauer$^7$ and revived by Wigner.$^{21}$  I shall explore an approach which
seeks the explanation of the reduction of the wave packet in the fact that
the apparatus, and, for that matter, any ``classical'' system is open, not
closed.  The goal of the quantum theory of measurement---accounting for the
transition $\rho_{pure} \rightarrow \rho_{mix}$, Eq. (2.7)---is realized by
the interaction of the apparatus with the environment. As a consequence
correlation term of the pure state density matrix are damped away.  Thus,
the system cannot be observed in a superposition of the states that appear
on the diagonal of $\rho_{mix}$:  This can be interpreted by saying that
there are effective superselection rules induced by the
apparatus-environment information transfer.

\noindent
{\bf III. \hspace {.15 in} ENVIRONMENT INDUCED SUPERSELECTION RULES}

The purpose of this section is to demonstrate on an exactly soluble example
how the interaction of a quantum system with its environment may induce
superselection rules.  I shall begin by investigating an interaction of a
pair of two state systems---a bit by bit ``measurement''.  Next I shall
demonstrate how a single atom ``environment'', performing a non-demolition
measurement on the apparatus
atom can bring about the second stage of the measurement process.  Finally,
I shall show how the interaction with the many atom environment can induce
superselection rules and single out the preferred pointer basis.  This same
mechanism for inducing superselection rules can operate in ``real world''
situations.
\newpage
\noindent
\underline{Bit-by-Bit ``Measurement"}

Consider a pair of two-state systems. I shall call one of them ``spin'' and
the other of them ``atom''.  Basis states of the spin can be denoted by
$|\uparrow\rangle$, $|\downarrow\rangle$.  Alternative basis can then be
written
as:
$$
|\odot\rangle = \left(|\uparrow\rangle + | \downarrow\rangle \right) /
\sqrt{2}
\ \ ,
\eqno(3.1a)
$$
$$
|\otimes\rangle = \left(|\uparrow\rangle - | \downarrow\rangle \right) /
\sqrt{2} \ \ ,
\eqno(3.1b)
$$
or;
$$
|\rightarrow\rangle = \left(|\uparrow\rangle + i| \downarrow\rangle \right) /
\sqrt{2} \ \ ,
\eqno(3.2a)
$$
$$
|\leftarrow\rangle = \left(|\uparrow\rangle - i| \downarrow\rangle \right) /
\sqrt{2} \ \ ,
\eqno(3.2b)
$$
Basis states of the atom consist of the ``ground state'' $|\mp\rangle$ and the
``excited state'' $|\pm\rangle$.  Despite this nomenclature, I shall assume
that
the energy of the ``atom'' in either of these states is identical.
Alternative basis states of the atom are then:
$$
| +\rangle = \left(|\pm\rangle + | \mp\rangle \right / \sqrt{2} \ \ ,
\eqno(3.3a)
$$
$$
| -\rangle = \left(|\pm\rangle - | \mp\rangle \right / \sqrt{2} \ \ ,
\eqno(3.3b)
$$
and;
$$
| _\top\rangle = \left(|\pm\rangle + i| \mp\rangle \right / \sqrt{2} \ \ ,
\eqno(3.4a)
$$
$$
| _\bot\rangle = \left(|\pm\rangle - i| \mp\rangle \right / \sqrt{2} \ \ ,
\eqno(3.4b)
$$

To investigate the process of the transfer of information consider an
interaction between the atom and the spin given by the hamiltonian:
$$
H^{AS} = g\left(| _\bot\rangle \langle  _\bot| - | _\top\rangle\langle _\top|
\right) \otimes
\left(|\uparrow\rangle\langle\uparrow| -
|\downarrow\rangle\langle\downarrow|\right)
$$
$$
\hspace {.15 in} = ig \left(|\pm\rangle\langle\mp| - |
\mp\rangle\langle\pm|\right) \otimes \left(
|\uparrow\rangle\langle\uparrow| - |\downarrow\rangle\langle\downarrow|
\right)
\ \ . \eqno(3.5)
$$
Suppose that the initial state of the atom is $|+\rangle$ and the initial
state
of
the spin is $a|\uparrow\rangle + b|\downarrow\rangle$:  hence the combined
state
vector
$|\Phi\rangle$ is initially given by the direct product:
$$
|\Phi_i\rangle = |\Phi (0)\rangle = | +\rangle \otimes \left(a |
\uparrow\rangle
+ b |
\downarrow\rangle \right) \ \ .  \eqno(3.6)
$$
As a result of the hamiltonian evolution generated by $H^{AS} \ |\Phi\rangle$ is
transformed into:
$$
|\Phi(t)\rangle = a|\uparrow\rangle \otimes \left[\sin (\pi/4 + gt) |
\pm\rangle
+ \cos
(\pi/4 + gt) |\mp\rangle \right]
$$
$$
+ b|\downarrow\rangle \otimes \left[ \sin(\pi/4 - gt) | \pm\rangle +
\cos(\pi/4
- gt)
| \mp \right \rangle ] \ \ . \eqno(3.7)
$$
Above, and throughout most of this paper I shall set; $\hbar  =
1$.  Consider now the state of the combined atom-spin system when $t =
\pi/4g = \tau$.  Then:
$$
|\Phi_f\rangle = |\Phi(\tau)\rangle = \left(a|\uparrow\rangle \otimes |
\pm\rangle + b|\downarrow\rangle
\otimes | \mp\rangle \right) / \sqrt{2} \ \ . \eqno (3.8)
$$
This is still, beyond any doubt, a pure state.  However, both the spin and
the atom, when treated as individual systems, are in mixed states.  This
can be easily shown by calculating their density matrices:
$$
\rho^S = | a |^2 | \uparrow\rangle \langle \uparrow | + | b |^2 |
\downarrow\rangle \langle
\downarrow | \ \ , \eqno(3.9a)
$$
$$
\rho^A = | a |^2 | \pm\rangle \langle \pm | + | b |^2 | \mp\rangle \langle
\mp |
\ \ . \eqno(3.9b)
$$
The off-diagonal terms vanish.  This accomplishes the apparent collapse for
each of the subsystems described by the total, still pure wave function
$|\Phi(\tau)\rangle $.\footnote{A thought experiment which realizes the
transition from the initial $|\Phi(0)\rangle$ given by Eq. (3.6) to the
final $|
\Phi (\tau)\rangle$ of Eq. (3.8) employing the reversible Stern-Gerlach
apparatus
has been recently described by Scully, Shea and McCullum$^{22}$ and
Scully$^{23}$.}

\noindent
\underline{Environment-Induced Superselection Rules}

Having concluded the first stage of the measurement---an establishment of
the correlation between the apparatus-atom and the spin-system, Eq.
(3.8)---one may ask if there is any reason to worry about the reduction of
the state vector.  After all, perfect correlation implies that by
consulting the atom and ``reading off'' its state (either $|\pm\rangle$ or
$|\mp\rangle$ ) one can find out whether the spin is $|\uparrow\rangle$ or
$\downarrow\rangle$--- parallel or antiparallel to the Z-axis.  What more can
one
ask from a perfect measeurement?  To expose the difficulty associated with
the purity of $|\Phi(\tau)\rangle$ let us rewrite Eq. (3.8) using $|+\rangle$,
$|-\rangle$ as
a basis for the atom Hilbert space.  For simplicity, we consider the case
when $a = b = 1/\sqrt{2}$.  Then:
$$
|\Phi(\tau)\rangle = \left(|\uparrow\rangle \otimes| \pm\rangle +
|\downarrow\rangle \otimes |
\mp\rangle \right) / \sqrt{2}
$$
$$
\hspace{.55 in} = \left(|\odot\rangle \otimes | +\rangle + | \otimes\rangle
\otimes | -\rangle
\right) / \sqrt{2} \ \ .  \eqno(3.9)
$$
Thus, by ``looking'' at the apparatus in the different basis one can find
out whether the spin is parallel or antiparallel to the X-axis!  And the
decision on how (in what basis) shall one measure the state of the atom
and, hence, in what direction will the spin eventually point can be made
long after the spin-atom interaction has ceased!  This affords another
example of the counterintiutive ``delayed-choice'' features of quantum
phenomena, first discussed by Wheeler$^{24, 25}$ and intimately connected
with the so-called ``nonseparability''.  Certainly, no real world measuring
apparatus leaves this delayed-choice option open for the experimenter.  And
the purity of $|\Phi (\tau)\rangle$ is the origin of this surprising
feature.\footnote{The relation between the nonseparability, complemntarity
and delayed-choice was further discussed on the example of Einstein's
version of double slit experiment by Wootters and Zurek$^{26}$.  A practical
realization of such experiments has been discussed by Wickes, Alley and
Jakubowicz as well as by Bartell$^{27}$.}

Information transfer from the systems (atom, spin) into correlations
between them was the ultimate source of the apparent collapse of the state
vector of both spin and atom.  Can this mechanism of information transfer
be employed to bring about the transition from $\rho_{pure}$ to
$\rho_{mix}$ for the combined apparatus-system object?  To see how, let us
begin with the pure wave-function $|\Phi (\tau)\rangle$, given by Eq. (3.8).
Let
us suppose that the atom plays the role of the apparatus and the spin is
the measured system.  To remove the off-diagonal terms from the matrix;
$$
\rho_{pure}^{AS} = |\Phi (\tau)\rangle \langle \Phi (\tau) | = |a|^2
|\pm\rangle
\langle \pm |
\otimes| \uparrow\rangle \langle \uparrow | + |b|^2 |\mp\rangle \langle \mp |
\otimes |
\downarrow\rangle \langle \downarrow |
$$
$$
+ ab* |\pm\rangle \langle \mp | \otimes | \uparrow\rangle \langle
\downarrow | +
a*b | \mp\rangle \langle \pm
| \otimes | \downarrow\rangle \langle \uparrow | \ \ ,  \eqno(3.10)
$$
we can couple the apparatus-atom with the two-state ``environment
atom''.  The states of the environment atom are designated by parentheses
rather than Dirac's brackets; $| \pm )$, $|\mp)$; $|+ )$, $|- )$;
$| _\bot)$, $| _\top )$, respectively.  The coupling hamiltonian is:
$$
H^{AE} = g_1 (| _\bot) ( _\bot| - | _\top) ( _\top |) \otimes (|\pm\rangle
\langle
\pm | - |\mp\rangle \langle \mp|)
$$
$$
\hspace{.55 in} = ig_1 (|\pm) (\mp| - |\mp) ( \pm |) \otimes (|\pm\rangle
\langle \pm
| - |\mp\rangle \langle \mp|) \ \ . \eqno(3.11)
$$
When the initial state of the environment is $|+)$, i.e., the initial state
of the combined apparatus-system environment is:
$$
|\psi_i\rangle = (a |\uparrow \rangle \otimes | \pm \rangle + b|\downarrow
\otimes | \mp \rangle ) \otimes | + \rangle \ \ .  \eqno(3.12)
$$
Then the final state for $g_1t = \pi/4$ will be:
$$
|\psi_f\rangle = (a |\uparrow \rangle \otimes | \pm \rangle \otimes| \pm ) +
b|\downarrow \otimes | \mp \rangle  \otimes | + \rangle ) \ \ .  \eqno(3.13)
$$
When the states of the environment atom are traced out, one obtains a final
state density matrix:
$$
\rho_{mix}^{AS} = |a|^2 |\uparrow \rangle \langle \uparrow | \otimes | \pm
\rangle \langle \pm | + |b|^2 |\downarrow \rangle \langle \downarrow |
\otimes |
\mp \rangle \langle \mp | \eqno(3.14)
$$

This shows how, employing partial information transfer from the apparatus
to the
environment, one can accomplish the transition $\rho_{pure} \rightarrow
\rho_{mix}$ required for the collapse of the state vector, Eq. (2.7).

It is important to distinguish essential and inessential requirements for
the occurrence of the environment induced collapse. One may, for instance,
object that for the real world environments conditions of terminating the
interaction at $g_1t = \pi/4$ or starting from a particular, pure initial
state are artificial and hence the proposed solution is not applicable to
the measurement problem.  To counter these objections I shall investigate a
more realistic case when the environment consists of many noninteracting
atoms, each of them coupled to the apparatus atom by the hamiltonian
$H^{AE}_k$:
$$
H^{AE}_k = ig_k (|\pm)(\mp| - |\mp)(\pm|)_k \otimes (|\pm \rangle \langle
\pm | - |\mp \rangle \langle \mp | ) \prod_{j \neq k} \otimes 1_j
$$
$$
= -g_k (| _\bot)( _\bot| - | _\top) ( _\top|)_k \otimes (|\pm \rangle
\langle \pm | - | \mp \rangle \langle \mp | ) \prod_{j \neq k} \otimes 1_j
\eqno(3.15)
$$
then the resulting wave function $j \neq k$, at an arbitrary time $t$, is:
$$
|\psi(t) \rangle = a| \uparrow \rangle \otimes | \pm \rangle
\prod_{k=1}^{N} \otimes \left[\alpha_{k\uparrow} (t) | _\bot) +
\beta_{k\uparrow} (t) | _\top) \right]
$$
$$
  + b|\downarrow \rangle \otimes | \mp \rangle \prod_{k=1}^{N} \otimes
\left [ \alpha_{k\downarrow} (t) | _\bot) + \beta_{k\downarrow} (t)
| _\top) \right ]  \eqno(3.16)
$$
Here $N$ is the total number of the environment atoms, and:
$$
\alpha_{k\uparrow} (t) = \alpha_k \exp (-ig_kt) ; \ \ \alpha_{k\downarrow}
(t) = \alpha_k \exp (+ig_kt)
$$
$$
\beta_{k\uparrow} (t) = \beta_k \exp (+ig_kt) ; \ \ \beta_{k\downarrow} (t)
= \beta_k \exp (-ig_kt)
$$

Constants $\alpha_k$ and $\beta_k$ are fixed by the initial states of the
environment atoms.  Now one can calculate the density matrix describing the
system-apparatus combination by taking the partial trace over the states of
the environment:
$$
\rho^{AS} = |a|^2 |\uparrow \rangle \langle \uparrow | \otimes |\pm \rangle
\langle \pm | + |b|^2 |\downarrow \rangle \langle \downarrow | \otimes \mp
\rangle \langle \mp |
$$
$$
+ z(t)ab^*| \uparrow \rangle \langle \downarrow | \otimes | \pm \rangle
\langle \mp | + z^*(t) a^*b |\downarrow \rangle \langle \uparrow | \otimes |
\mp \rangle \langle \pm | \ \ . \eqno(3.17)
$$
Above the correlation-damping factor $z(t)$ is given by:
$$
z(t) = \prod_{k=1}^{N} \left [ \cos 2g_kt + i \left(|\alpha_k|^2 -
|\beta_k|^2 \right) \sin 2g_kt \right] \eqno(3.18)
$$
When the absolute value of the correlation-damping factor $z(t)$ is much
less than unity, the off-diagonal terms of the density matrix effectively
disappear. The correlations between the apparatus and the measured system
are, nevertheless, left intact in the preferred pointer basis $\left[|\pm
\rangle, \ |\mp \rangle \right]$ of the apparatus.

\noindent
\underline{Information Conservation and the Irreversible Decay of Correlations}

It is not difficult to show, that $z(t)$ starting initially $(t = 0)$ at
$z(0) = 1$, will quickly fall off to values $z(t) \simeq  0$.  In particular,
when the coupling constants $\{g_k\}$ are chosen at random one can readily
show that:

(i) \hspace{.15 in} $ \langle z(t)\rangle = 0$ ;

(ii) \hspace {.15 in} $\langle |z(t)|^2 \rangle = \prod_{k=1}^{N} \left [
(1 + \gamma_k^2)/2 \right ]$ ;

\hspace {.15 in} where $\gamma_k = \left | |\alpha_k|^2 - |\beta_k|^2
\right |$ .

\noindent
Thus, unless $\gamma_k = 1$ for all $k$ (the environment atoms are in the
eigenstates of the apparatus-environment interaction hamiltonian) the
information transfer from the apparatus to the environment will cause the
effective collapse since:
$$
\left( \langle |a(t)|^2 \rangle \right) ^{\frac{1}{2}} < < 1 \ \ . \eqno(3.19)
$$
A typical plot of $z(t)$ for three environments of increasing size $(N = 5,
10, 15)$ is shown in Figure 1.  It is apparent there that even a relatively
small environment of $N = 15$ is capable of making $z(t) \simeq 0$ for a
very long time.  It is, however, also clear that the total state of the
combined, apparatus-system-environment object evolves in a unitary fashion,
and no information is really irreversibly destroyed.  The decay of $z(t)$
is due to the transfer of information, and not due to the destruction of
information.  Thus, if the environment of the real world apparata were an
isolated, closed system, then after a sufficiently long period of time $T$
the value of $z(t)$ would have to return back arbitrarily close to the
initial $z = 1$. In particular, when all $g_k$ are commensurable, $z = 1$
would recur exactly.  The recurrence of correlations encountered here is
closely reminiscent of the Poincar\'{e} recurrence cycle of classical
mechanics.$^{41}$

\begin{figure}
\vspace{0.5in}
{\bf THIS FIGURE IS ESSENTIALLY IDENTICAL WITH THE ILLUSTRATION OF THE
CORRELATION-DAMPING FACTOR IN} W. H. Zurek, {\it Environment - Induced
Superselection Rules}, Physical Review {\bf D26}, 1862 (1982). 
{\bf THE ORIGINAL CAN BE OF COURSE FOUND IN THE PROCEEDINGS, 
SEE THE FOOTNOTE ON THE VERY FIRST PAGE OF THIS PAPER}

\caption{Correlation-damping factor $z(t)$, see Eq. (3.17) and Eq. (3.18),
for the spin in a two-state atom environment. Number of the environment
atoms is, respectively, (a) $N = 5$; (b) $N = 10$; (c) $N = 15$.  Atom-spin
coupling constants $g_k$ of Eq. (3.18) were chosen randomly with the
distribution uniform in the open interval $(0,1)$.  For simplicity the
initial conditions $|\alpha_k| = |\beta_k|$ was imposed for all atoms, and,
consequently, $z(t)$ is purely real.  Note disappearance of the recurrences
with the increase of $N$.}
\end{figure}

One could expect, that the ultimate ``decay of correlations'' and the
disappearance of recurrences could take place in the limit of large
$N$.  This reasoning, often adopted in the discussions of irreversibility
and quantum measurements$^{28-32}$ contains at best only a part of the
collapse mechanism:  it shows that in the hamiltonian evolution information
can be transferred into those degrees of freedom which are deemed ``not
observable'' due to some external
restriction.  The rationale behind this procedure is very similar to the
old and well-known idea of ``coarse graining''.  Taking limits $\tau
\rightarrow \infty$ or $N \rightarrow \infty$ may help in finding out what
are the macroscopic degrees of freedom, but provides no real clue as to why
there is a collapse of the wave function; the \underline{total} wave
function is still pure.  Moreover, neither of these limits is relevant in
the case of real, and hence finite in size and duration, physical
experiments.\footnote{This point was expressed particularly clearly by
Bell$^{33}$ in his discussion of Hepp-Coleman model$^{30}$.} As long as the
von Neumann (Schr\"{o}inger) equation remains rigorously valid, there can
be no rigorous collapse of the wave packet of the closed system.

The situation is quite different for open systems.  When the information is
transferred from the system to the environment, and the environment itself
is also an open system---as in practice is always the case---then the
information is ``passed on'' and continues spreading away from where it has
originated.  Evolution of the open system is no longer unitary.  The
effective collapse takes place not because of the thermodynamic or ergodic
properties of the investigated system:  it is rather a consequence of its
interaction with the environment.  Part of this idea is contained already
in the ``usual'' derivations of the master equations for quantum systems by
means of projection operators$^{34}$.  There the role of the environment is
played by the correlation degrees of freedom.  Moreover, difficulties of
the usual derivation of an irreversible master equation, (see, for example,
a discussion of Zeh$^8$) e.g. the necessity of the repeated ``random phase
approximation'' can be resolved when the role of the environment is
properly taken into account:  the environment itself continuously
recollapses the state of the system.

\noindent
{\bf IV.  \hspace{.15 in} INFORMATION TRANSFER}

The aim of this section is to provide a mathematical definition of the
concept of transfer of information.  I shall follow closely Everett's
discussion$^{36}$ which employs Shannon's definition of
information$^{37}$.  For a density matrix $\rho$ in an N-dimensional Hilbert
space information $I$ is given by the formula:
$$
I= \ln N - Tr(\rho \ln \rho) \ \ . \eqno(4.1)
$$
It is well known that if $\rho$ evolves according to von Neumann equation $I$
is a constant of motion.  I shall use information defined by Eq. (4.l) in
the discussion of the information transfer between the spin and the atom in
the bit-by-bit ``measurement'' described in the previous section.  An
extrapolation of the formalism presented below to more general situations
is conceptually straightforward, even though it
does encounter technical difficulties in the case of continuously infinite
Hilbert spaces.  Some of these difficulties have been already pointed out
by Everett.$^{36}$

Consider the initial wave function $|\Phi(0) \rangle$ of the combined
spin-atom object,
given by the direct product, Eq. (3.6).
$$
|\Phi(0) \rangle = | + \rangle \otimes (a|\uparrow \rangle + b| \downarrow
\rangle ) \ \ .
$$
The total system as well as each of the subsystems is in a pure state.  The
total information can be then readily evaluated:
$$
I^{AS} = \ln 4 - Tr \rho^{AS} \ln \rho^{AS} = \ln 4 \eqno(4.2)
$$
This total information can be expressed as a sum of two separate
contributions, originating in two subsystems, the atom and the spin:
$$
I_A = \ln 2 - Tr \rho^A \ln \rho^A = \ln 2 \eqno(4.3a)
$$
$$
I_S = \ln 2 - Tr \rho^S \ln \rho^S = \ln 2 \eqno(4.3b)
$$
Consider now the final wave function:
$$
|\Phi(\tau)\rangle = \left(a|\uparrow \rangle \otimes | \pm \rangle + b|
\downarrow \rangle \otimes | \mp \rangle \right)
$$
given by Eq. (3.8).  Since the combined system is still in a pure state,
the total information must be still $I^{AS} = \ln 4$, the same as before the
establishment of the correlation.  This is no longer true of the
subsystems.  The respective informations are, in general, less than $\ln 2$:
$$
I^A = I^S = \ln 2 - \left( |a|^2 \ln|a|^2 + |b|^2\ln|b|^2 \right) \ \ .
\eqno(4.4)
$$
In particular, when $a = b = 1/\sqrt{2}$, $I^A = I^S = 0$.  In the generic
case
$$
{\mathcal{I}} = I^A + I^S < I^{AS} \ \ . \eqno(4.5)
$$
Where is the missing information $I_c$?
$$
I_c = I^{AS} - {\mathcal{I}} \eqno(4.6)
$$
It does not take long to realize that $I_c$ resides in the correlations
between the systems, and not in the systems.

In terms of information transfer the bit-by-bit ``measurement'' in the
particular case $a = b = 1/\sqrt{2}$ can be described by stating that the
definiteness of the states of the separate systems has been transformed
into the definiteness of the correlations between them.  Or, to put it
still differently, before the interaction between the spin and the atom
their correlations with the observer (who has set up their initial states)
were definite:  After the interaction the correlations between the observer
and the two subsystems disappear; they are replaced by the correlations
between the subsystems.  Of course, no information is lost:  it is only
transferred from one set of correlations to the other.

One can generalize the formalism of information transfer to arbitrary
systems in finite-dimensional Hilbert spaces.  Suppose the combined system
under consideration consists of $N$ subsystems, the k$^{th}$ subsystem
being described by the $n_k$ dimensional Hilbert space
${\mathcal{H}}_k$.  The total Hilbert space is then given by the direct
product:
$$
{\mathcal{H}} = \prod^{N}_{k=1} \otimes {\mathcal{H}}_k \ \ . \eqno(4.7)
$$
The total information $I$---a constant of motion under the evolution
generated by the von Neumann equation---is given by:
$$
I = \ln \left(\prod_{k=1}^{N} n_k \right) - Tr (\rho \ln \rho) \ \ . \eqno(4.8)
$$
where $\rho$ is the density  matrix of the whole system.  The information
residing in the k$^{th}$ subsystem is given by:
$$
I_k = \ln(n_k) - Tr \rho_k \ln \rho_k \eqno(4.9)
$$
where $\rho_k$, density matrix of the k$^{th}$ sybsystem, is obtained from
$\rho$ by a partial trace.  The total information in the subsystems is
equal to:
$$
{\mathcal{I}} = \sum^{N}_{k=1} I_k \eqno(4.10)
$$
The total information in correlations between the subsystems is given by:
$$
I_c = I - {\mathcal{I}} = I - \sum^{N}_{k=1} I_k \eqno(4.11)
$$
When $I_c = 0$ the subsystems are uncorrelated.  In quantum measurements
this situation occurs at the beginning of stage 1, before the correlations
between the two subsystems of interest---the apparatus and the measured
system---are set up.  The first stage of measurement can be described as a
transfer of the information from the observer-apparatus correlations to the
apparatus-system correlations.  After the completion of this stage the
total information of the apparatus system combination is still the same as
it was before their interaction has started.  This excessive amount of
correlation is responsible for the paradoxes associated with
irreversibility and non-separability.  To prevent the occurrence of these
paradoxes von Neumann postulated the reduction of the state vector.  In the
course of this process the off-diagonal terms of the density matrix, which
contain information about spurious correlations, are supposed to
disappear.  The key difficulty of the quantum theory of measurement is the
incompatibility of the reduction postulate with the unitarity of the
evolution equations.  In other words, in a closed system it is impossible
to get rid of this spurious information.  The resolution considered here
asserts that the spurious information is removed by the environment.  Thus,
it is never actually destroyed: we ignore it, because it isn't a part of
our combined system anymore.

\noindent
{\bf V. \hspace{.15 in} DISCUSSION}

The aim of these lectures was to investigate the idea that the second stage
of the measurement process---the apparent reduction of the state
vector---can be accomplished through a transfer of a part of the
information from the apparatus-system into the apparatus-environment
correlations.  All the evolutions are unitary:  the information is
ultimately never destroyed---it is only displaced. The \underline{pointer
basis} of the quantum system is subjected to a continuous measurement
performed on it by the environment, which plays the role of an (unread)
apparatus.  As a result, effective \underline{superselection rules}
arise.  Continuous destruction of the phase information causes the pointer
observable to behave in a way which makes it appear as ``classical''.  It
is extremely important to stress that this classical behavior of the
pointer observable is not a consequence of just the structure of the Hilbert
space in which it is defined:  The exact pointer observable must commute
with the \underline{total} hamiltonian of the investigated system (i.e. the
self-hamiltonian and the system environment interaction
hamiltonian).  Classical properties of a quantum system are induced by the
``measurements'' performed on it by the environment.  In this sense one can
claim that properties of the system originate in part in its interactions
with the environment. The system is ``the way it is'' because it is
perceived by the environment in a particular way.  This conclusion is in
contradiction with intuitions based on classical phenomena, and is a direct
consequence of quantum mechanics.

Environment-induced superselection rules, counterintuitive as they may
seem, answer many of the questions concerning the apparent collapse of the
wave packet.  There are, of course, many more questions they allow to be
raised.  The aim of this section is to ask---and, at least in part,
answer---some of these questions.

\noindent
\underline{The Density Matrix in the Presence of the
Environment-Induced} \newline  \underline{Superselection Rules}

Bell$^{33}$ has once said that ``In quantum measurement theory elimination
of ... [the off-diagonal density matrix terms expressing] ...
coherence is the philosophers stone''.  Selective information transfer
resulting in the environment-induced superselection rules accomplishes this
task by transferring spurious correlations into the environment.  The first
question that one can raise is whether transfer of correlations
accomplishes the same goal as their destruction would.  I shall come back
to it later in this section.  Here let me consider a simpler, but equally
important problem and ask:  what have we really gained by destroying
correlations?  Clearly, predictive power of quantum theory has been
enhanced by our considerations, but only in a very special way:  one can
tell now what a given apparatus measures, but---as before---one cannot tell
what the outcome of the  measurement will be.  The diagonal part of the
density matrix is still the same, giving only probabilities of different
possible outcomes of the measurement process.  This is so despite the fact
that when someone consults the apparatus the outcome he is going to see
will be a definite eigenvalue of the measured operator---a definite
eigenstate of the pointer observable.  And we still do not know
\underline{which} eigenstate it is going to be.  We have, nevertheless,
learned into what mixture shall the combined apparatus-system state vector
collapse.  Hence we know what information about the system still remains in
the apparatus.

The aim of the discussion in this subsection is to explore the
interpretation according to which a density matrix diagonal in the pointer
basis describes, in presence of the environment-induced superselection
rules, a quantum system which is in a definite state---in one of the
pointer basis eigenstates.  The probabilistic character of the density
matrix results therefore, according to this point of view, solely from our
ignorance of the true state of the system---and not from the fact that the
system itself is not in a definite state.

This interpretation may seem obviously correct.  For, let me suppose that
my friend prepares an ensemble of spins, one by one, in states $|\uparrow
\rangle$ or $|\downarrow\rangle$ with respective probabilities
$\rho_\uparrow$ and $\rho_\downarrow$.  Then one can maintain that any
particular spin in this ensemble is in a well- defined state (i.e. my
friend knows that he has prepared spin \#137 in the state
$|\uparrow\rangle$, etc.).  However, if my friend refuses to share this
detailed information with me and tells me only values of $\rho_\uparrow$
and $\rho_\downarrow$, I shall have to trace out the memory of my friend
and represent each spin by a density matrix:
$$
\rho_{spin} = \rho_\uparrow |\uparrow\rangle \langle \uparrow | +
\rho_{\downarrow} |\downarrow \rangle \langle \downarrow| \ \ . \eqno(5.1)
$$
In the situation described above the density matrix is then nothing else
but the expression of my ignorance:  I do not know in which of the
eigenstates appearing on the diagonal $\rho_{spin}$ particular spin
(\#137) was prepared.

Can one always apply this interpretation to \underline{any} diagonal density
matrix?  Unfortunately not.  To see why, let me suppose that
$\rho_\uparrow$ >
$\rho_\downarrow$. Then $\rho_{spin}$ can be rewritten, for instance as:
$$
\rho_{spin} = (\rho_\uparrow - \rho_\downarrow) |\uparrow \rangle \langle
\uparrow | +
\rho_\downarrow (|\odot\rangle\langle\odot| +
|\otimes\rangle\langle\otimes|) \ \ .
\eqno(5.2)
$$
That is, my friend could have prepared the spin ensemble by mixing spins
in  the state $|\uparrow\rangle, |\odot \rangle$ and $|\otimes\rangle$ in
the appropriate proportions. In fact, there are continuously many ways in
which he could have prepared the ensemble described by Eq. (5.1).  And only
in one of these ways the initial claim---that the system is in a
well-defined state, either in $|\uparrow\rangle$ or in
$\downarrow\rangle$---has any justification.  In all the other instances,
as our friend can testify, this claim is simply wrong.

This well-known$^{38}$ nonuniqueness of the ensemble described by a given
density matrix disappears in presence of the environment- induced
superselection rules.  The mixture is now diagonal in the uniquely
specified pointer basis.  The environment---which continuously measures
systems under consider-ation---contains a record of their individual
states.  The troublesome Equation (5.2) can be of course written, but it
cannot by physically realized.  For, even if our friend will prepare some
of the spins in the state $|\odot\rangle$ or $|\otimes\rangle$, or, for
that matter, in any state other than one of the pointer basis states (which
we can assume for the sake of the argument to be $\{|\uparrow\rangle$,
$|\downarrow\rangle\}$), the state of each spin will almost immediately
``collapse'' to a mixture diagonal in the pointer basis.  That is:
$$
(a|\uparrow\rangle + b|\downarrow\rangle)(a^*\langle\uparrow| +
b^*\langle\downarrow|)
\rightarrow |a|^2|\uparrow\rangle\langle\uparrow| +
|b|^2|\downarrow\rangle\langle\downarrow|
\
\ .
$$
The record of our friend is no longer trustworthy---it is the environment
which keeps the true record of what the spin is.

Therefore, one must conclude that in absence of the superselection rules
the same density matrix describes, in the generic case, continuously many
different ensembles.  This nonuniqueness is removed by the
environment-induced superselection rules:  a density matrix diagonal in the
pointer basis describes now a unique ensemble consisting of individual
systems each of which can be thought of as being in the specific eigenstate
of the pointer observable.  This interpretation, tempting as it is, still
faces a major problem:  The environment has not destroyed the correlation
information, and so one can presumably still reverse the evolution of the
complete (i.e., including the environment) system to restore the initial
wavefunction.  I shall discuss this problem in the last subsection of this
section.

\noindent
\underline{Two Origins of the Entropy Increase}

According to von Neumann the second stage of the measurement process is
explicitly irreversible:  the information is destroyed. In the model of the
reduction of the state vector investigated here the information is only
displaced.  Can one consider this an irreversible process?  And---assuming
that the measurement theory proposed here reflects accurately the real act
of measurement---does it help to understand the second law of
thermodynamics?  It is not yet possible to give a final, complete answer to
these questions.  The following few remarks have therefore a rather
preliminary character.

Entropy and its definition is the content of the first remark. Let us
examine ${\mathcal{I}}$, the total information $I$ less the information
residing in the correlations $I_c$,
$$
{\mathcal{I}} = I - I_c
$$
as a candidate for the negative of the entropy (modulo a constant). For
simplicity, we shall call ${\mathcal{I}}$ ``negentropy''.  As defined in
the previous section, the negentropy ${\mathcal{I}}$ represents the
information available to the observer about the subsystems of the physical
system under investigation.  Once these subsystems---i.e. the partition of
the complete Hilbert space into subspaces--- is fixed, the negentropy
${\mathcal{I}}$ is a well-defined quantity.  It can be calculated directly
from Eq. (4.9), (4.10).  Of course, one may object to this definition of
entropy by arguing that it shifts the subjective element from the usual and
arbitrary ``coarse-graining'' to the equally arbitrary concept of the
subsystem.  This objection, correct as it is in classical statistical
mechanics cannot be simply transplanted into the context of quantum
mechanics.  For, in the classical phase space shifting from one
coarse-grained network of cells to the other can be achieved, in principle,
at no expense in terms of energy:  Classical mechanics does not recognize
that any information transfer can be always ultimately reduced to an
interaction.  In quantum mechanics, on the other hand, partition of the
original Hilbert space into smaller subspaces is no longer arbitrary:  It
depends on the structure of the coupling between the system and the
measuring apparatus, the pointer basis of the apparatus, as well as on the
hamiltonian of the investigated system.  Therefore, the change of the
concept of a subsystem is not just a purely ``mental'' act.  On the
contrary, it implies a change of the coupling hamiltonians and, presumably,
requires an expenditure of a certain amount of energy.  For example, in a
system of $N$ hydrogen molecules the natural ``coarse-graining'' will
recognize molecules, atoms and elementary particles.  The problem of
determining what are the natural subsystems of a certain system from the
complete hamiltonian is, of course, by no means trivial.  I shall not
attempt to solve it here.  Some light has been thrown on it by the idea of
subdynamics$^{39}$.  More recently, Deutsch$^{40}$ has considered the
product structure induced on the Hilbert space by the Hamiltonian.  His
approach, in a sense complementary to the one adopted here, gives also
equations for the preferred ``interpretation basis'', which is expected to
play a role similar to the pointer basis discussed in this paper.

The apparent increase of entropy is the subject of the second remark.  I
shall argue that its increase on the relaxation time scale is a consequence
of a mismatch between the division of a system into subsystems and the
evolution generated by the self-hamiltonian of the considered system, $H^S$
, which induces correlations between these subsystems.  Apart from this
short-term relaxation caused by the increase of the correlation information
$I_c$ on the expense of negentropy ${\mathcal{I}}$ , with $I$ constant, I
shall also consider in the last part of this subsection long-time
relaxation, occurring over periods of the order of the recurrence time of
the system, and caused by the interaction with the environment and
resulting in the decrease of $I$ itself.

Suppose that the total Hilbert space of the considered system is
partitioned into N subsystems, in a manner represented by a direct product,
Eq. (4.7).  I shall moreover suppose that one can measure arbitrarily
accurately any observable that is confined to one of the rigidly defined
$N$ subsystems, corresponding to $N$ subspaces ${\mathcal{H}}_k$.  As a
result of such an ideal measurement on the subspace ${\mathcal{H}}_k$ one
obtains an $\ell^{th}$ eigenvector $|k_\ell \rangle$ of the measured
observable $\hat{O}_k$.  After all the measurements are completed, the
state vector of the whole system is given by
$$
|s_{P_0}\rangle = |\Phi(0)\rangle = \prod_{k=1}^{N} \otimes|k_\ell\rangle \ \ .
\eqno(5.3)
$$
I shall call each state which can be expressed in the form of the above
product, with vectors $|k_\ell \rangle$ arbitrary but confined to the
sub-spaces
${\mathcal{H}}_k$ a ``product state''.  If the measurements within  each
${\mathcal{H}}_k$ are exhaustive (i.e. all the commuting observables  are
measured) then
the collection of all orthogonal products of the form of  Eq. (5.3)
provides a complete
basis in the space ${\mathcal{H}}$.  States  spanning this basis can be
numbered with an
index $P$, referring to the  permutation of the subscripts appearing on the
right hand
side of Eq.  (5.3).  Let us moreover suppose that the self-hamiltonian of
the system,
$H^S$, has a complete set of the eigenvectors $|\chi_j\rangle$:
$$
H^S|\chi_j\rangle = \epsilon_j|\chi_j\rangle  \eqno(5.4)
$$
In the generic case $H^S$ will be degenerate, i.e. not all $\epsilon_j$
will be different.  Now the general solution for the wave function which at
$t = 0$ was given by Eq. (5.3) can be written as:
$$
|\Phi(t)\rangle  =  \sum_{P,j} \left[\langle s_P|\chi_j\rangle \exp
(-i\epsilon_jt)\langle \chi_j|\Phi(0)\rangle\right]|s_P\rangle =
 \sum_{P} \alpha_{P_0P}(t)|s_P\rangle  \ \ . \eqno(5.5)
$$
Above evolutions can be classified into two categories, which I shall
tentatively call ``correlating'' and ``non-correlating''.  In the course of
a non-correlating evolution the state vector $|\Phi(t)\rangle$ can be for
all $t$ re-expressed as a product state:
$$
|\Phi(t)\rangle = \prod^{N}_{k=1} \otimes|k_\ell(t)\rangle \ \ . \eqno(5.6)
$$
That is, the self-hamiltonian causes state vectors of the subsystems to
rotate within their respective subspaces ${\mathcal{H}}_k$. This does not
necessarily mean that the subspaces ${\mathcal{H}}_k$ are
non-interacting:  It suffices for the states $|k_\ell\rangle$ appearing in
the product in Eq. (5.3) to be separately eigenstates of the hamiltonian
$H^S$.  For example, in the spin-atom interaction hamiltonian $H^{AS}$ Eq.
(3.5), states $| _\bot \rangle$, $|\downarrow \rangle$ had that property of
not becoming correlated.

Correlating evolution, on the other hand, introduces into
$|\Phi(\tau)\rangle$,  Eq. (5.5) terms which cannot be reduced to the form
of Eq. (5.6).  Evolution of the initial state vector $|\pm\rangle \otimes |
\odot \rangle$ under the influence of $H^{AS}$, Eq. (3.5) is an
example.  There are, moreover, interaction hamiltonians between the
subsystems which introduce correlations into any intial state vector
$|\Phi(0)\rangle$.  For such hamiltonians\footnote{An example of such
hamiltonians can be found in the lectures of Lugiato$^{42}$---it is used
there to justify the master equation.  See also discussion of measurement
by Stenholm$^{43}$ and his distinction between $T_1$ and $T_2$ processes.}
there is no eigenstate which can be decomposed into a product of the form
of Eq. (5.3). The increase of entropy is then caused by the increase of
correlations between the subsystems of the whole system.  Moreover, each
measurement brings the system into a product state with a product structure
conforming with the division into the subsystems.  Therefore, it decreases
the entropy of the system with respect to the measuring device by
destroying correlations between subsystems.  Consequently, following such a
measurement, one can expect that the natural evolution of the system will
result in a buildup of the correlations and in the increase of entropy.  It
is perhaps worth pointing out that the view adopted here is close to the
one of Boltzmann.  The key to the understanding of the entropy increase
lies, according to these proposals, in the distinction between the
information about the subsystems, which I have called negentropy
${\mathcal{I}}$, and the total information $I$, which is constant under the
unitary evolution.  Moreover, since measurements tend to always increase
${\mathcal{I}}$, by bringing the system closer to the product state
$|s_p\rangle$, and hence destroying the correlations, the evolution
following the measurement is likely to decrease ${\mathcal{I}}$.  An
important problem which has not been treated here in sufficient detail is
the problem of the division of the original system into the subsystems.  I
have assumed that such a division can be inferred for the system not
interacting with the environment from the total hamiltonian~ that is from
the $H^S$ plus the hamiltonian coupling the system with the measuring device.

When the system is not perfectly isolated, its entropy increases due to the
``leakage'' of the information into the environment.  This second process
leaves the system in a true mixture, which can be calculated by tracing out
states of the environment.  It might appear that this second process could
be made arbitrarily slow.  After all, there is nothing in principle to
prevent one from limiting, say, the rate of exchange of energy between the
system and its environment. The quantity that is of concern to us here is,
nevertheless, the information.  To conserve it one must keep the state of
the system, $|\Phi(t)\rangle$, uncorrelated with the state of the
environment.  That is, the total system-environment wave function
$|\Theta\rangle$ it must be true at all times that:
$$
|\Theta(t)\rangle = |\Phi(t)\rangle \otimes |\epsilon(t)\rangle \eqno(5.7)
$$
This may seem easy to achieve for the time intervals of the order of the
relaxation time of the system.  It is, however, doubtful whether such an
idealization can be valid for periods as long as the Poincar\'{e}
recurrence time.  This in turn implies that the difficulty with the quantum
version of Zermelo's paradox$^{41}$ disappears:  The state of the system
decays into a density matrix before it gets a chance to make a Zermelo
``comeback''.

To conclude this section let me restate its main points. According to the
view proposed here entropy can be regarded as a negative of the information
residing in the natural subsystems of the combined system.  This structure
is not arbitrary:  It is presumably rigidly defined by the self-hamiltonian
of the system with the measuring hamiltonian playing on additional
role.  The increase of entropy is caused by two distinct information
transfers.  First, there is a decrease of the negentropy ${\mathcal{I}}$
due to the buildup of the correlations.  It is responsible for the entropy
increase on the relaxation time scale.  Second, there is a much slower
buildup of the correlation between the system and its environment.  It is
responsible for the ultimate decay of the pure state into the density
matrix.  I believe that, combined, these two processes can explain both the
relaxation and the persistence of the thermodynamic equilibria.

\noindent
\underline{Amplification and Redundancy}

The aim of the following discussion is to argue that in the absence of an
exact pointer basis it is advantageous to make not one, but many copies of
the same information.  This can be achieved by using a collection of
microscopic systems, which together constitute a ``macroscopic''
apparatus.  The process of amplification facilitates the choice of the
reliable basis of the pointer, that is, that basis which keeps the
correlation with the measured system for a long time. (The exact pointer
basis retains the record for an arbitrarily long time).  This strategy of
making many copies of the same message is known in the communication
science as the strategy of using redundancy in order to increase
reliability of stored messages.

Consider a simple model for a macroscopic apparatus:  $N$ two-state
atoms.  Let the first stage of the measurement of a spin be represented by
a transition:
$$
	|\Phi_i\rangle  =  (a|\uparrow\rangle + b|\downarrow\rangle)
\prod^{N}_{k=1}
\otimes|+\rangle \longrightarrow
$$
$$
	\longrightarrow  a|\uparrow\rangle \prod^{N}_{k=1} \otimes | \pm
\rangle_k +
b|\downarrow\rangle \prod^{N}_{k=1} \otimes | \mp\rangle_k \ \
	 \eqno(5.8)
$$
In the absence of any particular symmetry in the apparatus-environment
interaction the two states; $\prod\otimes |\pm\rangle_k$, $\prod\otimes
|\mp\rangle_k$
are a natural choice for the reliable pointer basis of the  apparatus.  Of
course, these
two states do not span a complete basis in the
$2^N$ dimensional Hilbert space of the apparatus.  This is precisely the
meaning of redundancy: the space of possible massages is much larger than
the subspace actually employed to encode the result of the
measurement.  Now even in the presence of external perturbations there is a
clear-cut strategy which yields a reliable distinction between spin
$|\uparrow\rangle$ and $|\downarrow\rangle$.  Moreover, it is also possible
to distinguish between these two states of the spin by consulting only a
subset of all $N$ atoms.  To see why, consider a small apparatus,
consisting of just three recording atoms, and assume that the spin is
initially in a state $(|\uparrow \rangle + \downarrow \rangle)//2$. Then
the first stage of the measurement:
$$
|\Phi_i\rangle = |\odot\rangle\otimes| + \rangle\otimes| + \rangle\otimes|
+ \rangle
\longrightarrow
$$
$$
\longrightarrow  (|\uparrow\rangle\otimes| \pm\rangle\otimes| \pm\rangle
\otimes| \pm
\rangle + |\downarrow\rangle\otimes| \mp\rangle\otimes| \mp\rangle \otimes| \mp
\rangle)/(\sqrt{2})^3 = |\Phi_f\rangle
\eqno(5.9)
$$
Even if due to fluctuations or some inadequacies of the apparatus one
of  the atoms gets ``flipped'', the rule of a thumb ``the spin was $|\uparrow
\rangle$ if the majority of atoms was in the state $|\pm\rangle$'' suffices
to be correct. Moreover, if there are no disturbances and $|\Phi_f\rangle$ of
Eq. (5.8), (5.9) remains unperturbed, it is sufficient to consult only a
subset of $N$ atoms to learn about the state of the spin.  Of course, one
can still rewrite $|\Phi_f\rangle$ by using any alternative set of the
relative states.  For example:
$$
|\Phi_f\rangle =
\{|\uparrow\rangle\otimes[|\pm\rangle\otimes|\pm\rangle\otimes|\pm\rangle] +
|\downarrow\rangle \otimes
[|\mp\rangle\otimes|\mp\rangle\otimes|\mp\rangle]\}/(\sqrt{2})^3
$$
$$
\{|\odot\rangle\otimes [(|+\rangle\otimes|+\rangle\otimes|+\rangle) +
(|+\rangle\otimes|-\rangle\otimes|-\rangle) +
(|-\rangle\otimes|+\rangle\otimes|-\rangle) +
(|-\rangle\otimes|-\rangle\otimes|+\rangle)] +
$$
$$
|\otimes\rangle\otimes[(|-\rangle\otimes|-\rangle\otimes|-\rangle) +
(|-\rangle\otimes|+\rangle\otimes|+\rangle) +
(|+\rangle\otimes|-\rangle\otimes|+\rangle) +
(|+\rangle\otimes|+\rangle\otimes|-\rangle) ]\} /(\sqrt{2})^3
$$
Now an even number of atoms in the state $|-\rangle$ implies relative
state $|\odot\rangle$ while an odd number of $|-\rangle$ atoms imples the
alternative $|\otimes\rangle$.  This rule generalizes for the $\{|\odot
\rangle,$ $|\otimes\rangle\}$ spin basis to an arbitrary $N$.  Thus, it is
not difficult to see why, while there is nothing in principle which could
keep one from distinguishing between any pair of the relative spin states,
it is still easiest to tell $|\uparrow\rangle$ from
$|\downarrow\rangle$.  In particular, it will become successively more
difficult to distinguish between the states $|\odot\rangle$ and
$|\otimes\rangle$ and $N$ becomes truly macroscopic, $N \sim 10^{23}$ atoms
or so.

Redundancy is the reason why amplification is a good strategy for making
reliable records.  The reason why certain states seem a natural choice goes
beyond redundancy:  It contains also a basic assumption that the readout of
the information contained in the apparatus will recognize atoms as distinct
subsystems.  Consequently, if the information contained in some part of the
apparatus were to become obliterated, the remaining record would still
suffice to distinguish between $|\uparrow\rangle$ and $|\downarrow\rangle$
states of the spin.  This point of view in which apparatus is represented
by a collection of distinct Hilbert spaces has been discussed, for
instance, by Hepp$^{30}$, and more recently by Machida and Namiki$^{44}$,
and Araki$^{45}$.

\noindent
\underline{Pointer Basis and the Operational Interpretation of Quantum Theory}

The aim of this last subsection is to consider the relevance of the
environ-ment-induced superselection rules in the context of the
interpretation of quantum mechanics.  The emergence of the pointer basis
indicates, I shall argue, that from the standpoint of the observer external
to the system-apparatus object, the measurement yields a definite result
for a definite observable only when, apart from the system and
the apparatus, there is a third object, called here the
environment.  Presence of the environment is essential in making the choice
between many to-be-measured observables.

In discussions of quantum measurements in which the apparatus is explicitly
quantum (e.g., von Neumann's treatment) it was always tacitly taken for
granted that the quantum measuring apparatus somehow contains the
information determining which observable is going to be recorded.  This
distincion between various measuring procedures, so essential for Bohr's
concept of complementarity and in the Copenhagen interpretation of quantum
mechanics is not present for a truly quantum apparatus.  To see why we
return to the bit-by-bit ``measurement'' in the case of the perfect
nonseparable correlation, Eq. (3.8a).  There the apparatus-atom can always
supply a definite and exact value of the angular momentum of the spin in
any direction.  Thus, one is forced to admit that the atom ``knows'' the
component of the spin in an arbitrary direction with perfect
accuracy.  When the atom is considered to be a representative model of an
apparatus (in disagreement with the Copenhagen interpretation of quantum
theory) this last conclusion appears to be in a flagrant violation of the
principle of complementarity:  The apparatus-atom ``knows'' all about
complementary components of the spin.  Short of Copenhagen or Many World
interpretations (to which we shall come back) there is only one way of
avoiding a conflict between complementarity and nonseparability:  One must
recognize that while the two nonseparably correlated quantum systems have a
complete information about each other, they contain no information about
the outside Universe.

In particular, atom doesn't ``know'' what is the spin observable it has
recorded.  This information shall be provided by the correlation it will
establish with some other system---with the environment.  That is, the atom
does know all that is to be known about the state of the spin, but this
information cannot be expressed in terms of some absolute coordinate
system.  To see how the state of the measured system can be definite with
respect to the apparatus, despite the fact that the system-apparatus
combination remains in the state indefinite from the point of view of the
external observer, let us investigate a system composed of two spin
$\frac{1}{2}$ objects, described by the state vector:
$$
|\sigma\rangle = (|\uparrow\rangle_1 \otimes |\downarrow\rangle_2 -
|\downarrow\rangle_1
\otimes |\uparrow\rangle_2)/\sqrt{2} \eqno(5.10)
$$
Nothing can keep one from thinking of the first spin as of the measured
system and of the second spin as of a quantum apparatus.  After the first
stage of the measurement, which has resulted in the wave function
$|\sigma\rangle$, the state of the spin-system with respect to the spin-
apparatus is quite definite:  The (no. 1) spin-system always points in the
direction which is opposite to the orientation of the (no. 2)
spin-apparatus.  This is a definite, ``coordinate-independent''
statement.  Thus, from the point of view of the spin-apparatus, the
measurement has already yielded a definite result.  However, from the point
of view of an external observer the measurement has not been
completed:  The state of the spin remains indefinite with respect to the
state of the coordinate system in which this observer describes
physics.  The result of the spin-spin measurement, already definite from
the viewpoint of the spin-apparatus shall acquire definiteness with respect
to the external observer only when the apparatus and the observer will
become correlated.  Two distinct possibilities arise here.  First, the
observer may consult the apparatus.  Then the system-apparatus combination
will enter into a state that is definite with respect to this observer,
and, moreover, the observer will become aware of the outcome of the
measurement.  The second possibility may
arise when the apparatus enters into a correlation with the
environment.  Then the state of the apparatus with respect to the
environment becomes definite, and if the environment remains in a definite
state with respect to the observer, the state of the apparatus may as well
become definite with respect to the observer.  This does not imply that the
observer knows the outcome (see discussion in the first subsection of this
section).

\newpage
The key idea of the proposed above reinterpretation of quantum mechanics is
the \underline{operational} definition of the outcome of measurement.  The
states of any physical system are defined only with respect to the other
systems with which the system in question is actually correlated.  In this
sense operational interpretation of quantum mechanics is very close to the
Relative State interpretation of Everett.  \underline{The state of the
system is defined with respect to the} \underline{state of the observer.}
The new
element, present here but not in the original Everett's proposal is that
the distinct states of the observer should not be under certain
circumstances thought of as distinguishable by this observer.  The
motivation for such a reinterpretation of quantum theory comes not from the
dissatisfaction with the stochastic nature of quantum predictions.  Rather,
it arises from the following dilemma:  The observer (apparatus) presumably
``knows'' the outcome of the measurement as soon as the measurement is
performed.  Yet, this definiteness of the outcome is in conflict with the
principle of superposition, as it can be manifested by the reversibility of
the hamiltonian evolution:  Some ``higher order'' observer, for which the
first observer (an apparatus, Schr\"{o}dinger's cat, Wigner's friend,
$\cdots$) is just a term in the wave function can presumably consult the
first observer and learn this way about some definite outcome of the
measurement, or can reverse the evolution of the combined apparatus-system
object.  Such time reversal can bring back the initial state only if the
complete superposition, containing all the potential outcomes of the
experiment that could have been registered by the apparatus is still
present when the reversal procedure is initiated.  Thus, on one hand, the
apparatus is supposed to learn about the definite outcome, while on the
other hand it is supposed to be forced into a superposition corresponding
to all the conceivable outcomes.  Can one resolve this contradiction?  The
Copenhagen interpretation avoids it by dividing systems present in the
Universe into two distinct classes; microscopic, and therefore quantum, and
macroscopic, and therefore classical.  The Relative State (also known as
the ``Many World'') interpretation bypasses the problem by saying that the
second observer ``splits'' already when the first one splits.  Is there any
other, less drastic solution of this dilemma?  The only other proposal not
yet explored is the extension of the approach discussed above on the
example of two correlated spins.  There the outcomes that were distinct
from the standpoint of the external observer were indistinguishable to the
first ``observer''; spin-apparatus.  Can this explanation be employed to
resolve all the apparent paradoxes arising in quantum theory of measurements?

Our discussion certainly does not go far enough to yield a definite answer
to this question.  It may, in particular, be possible that the operational
point of view, in which all the observables are defined ``with respect to''
rather than in absolute terms will lead to predictions testably different
from those of Copenhagen interpretation. Let me, moreover, stress that the
operational interpretation of quantum measurements is close to the position
adopted by Bohr, who insists that quantum theory can be applied only by
macroscopic observers using macroscopic apparata.  Macrophysics is regarded
by Bohr as more fundamental than microphysics, for it is within the realm
of classical physics where ``a discrimination between different
experimental procedures which allows for the unambiguous use of the
complementary classical concepts''\footnote{This quotation is taken from
Bohr's reply$^{46}$ to the Einstein, Podolsky and Rosen argument.$^{4}$}
occurs.  In the operational approach to the process of measurement
similarly as in the Copenhagen interpretation it is the task of the
experimental arrangement to define what are the distinct recorded
properties.  Only after the to-be-measured property has been operationally
defined (e.g. by the direction of the field gradient in the Stern-Gerlach
apparatus) can a quantum phenomenon be brought to closure:  Via collapse of
the wavepacket the quantum system shall assume a definite eigenstate of the
measured observable. It is, however, absolutely crucial that this
observable, the ``property'' must be first, prior to the collapse, defined
by the experimental arrangement.  In this sense the property does not
belong entirely to the measured quantum system;  it is rather forced upon
it by the act of measurement.  The key objection against the Copenhagen
interpretation comes from the fact that it demands a parallel existence of
two distinct domains:  macro and microscopic.  This objection would not
apply to the operational interpretation.  (Of course there may be many,
much more serious objections which I have not yet discovered.)  Operational
interpretation attempts to give a unified point of view, where the whole
Universe is described by the same theory.  It is worth noting that the same
goal of a unified description of physics has been adopted by the Many World
(alias Many Universes, alias Relative State)
interpretation.$^{16,36,40}$  There, each time an interaction occurs
between two subsystems of the Universe, the whole Universe branches out
into as many new Universes as there are outcomes of the measurement.  The
key reason for this splitting comes from the assumption of existence of
abstract observables, which do not have to be realized operationally by
means of an experimental arrangement.  This point of view encounters
several difficulties. First of all, this splitting is unobservable, and
hence unphysical (but see reference$^{40}$).  Moreover, it is difficult to
accept the idea that our consciousness splits due to a collision of two
electrons in some yet-to-be discovered quasar on the outskirts of the
Universe. Furthermore, there are often difficulties in determining branches
into which the Universe should split.  This ambiguity is due to the freedom
of choice of the two relative bases.$^{40}$  Also, quite often the Universes
in which the splitting occurs must recombine to accommodate interference
experiments.  And last, but not least, adopting Many World interpretation
has not led to any new results in, for example, quantization of General
Relativity, as it was originally anticipated.$^{16,36}$ Above brief
discussion can do justice neither to the Many World interpretation nor to
the difficulties it encounters. Nevertheless one is led to believe that,
like the Copenhagen interpretation the Many World interpretation conceals
rather than explains problems inherent in quantum formalism.

The operational interpretation explored preliminarily above asserts that
there is no wave function for the whole Universe.  Every observer has his
own description (wave-function, or, more generally, density matrix) of the
physical systems around him.  This description is always expressed in a
language appropriate for that observer, e.g. the measuring spin
$\frac{1}{2}$ particle describes the rest of the world (all he can see)
like this: $|$anti-aligned with me$\rangle$.  It is quite conceivable that
one could develop this very preliminary proposal into a consistent
interpretation of quantum theory.  It is also possible, that by insisting
on the observables that are defined only ``with respect to'' rather than in
some absolute sense one may arrive at a theory that is significantly
(testably) different from quantum mechanics.  Preliminary investigation of
the basic ideas of the operational interpretation does not yet allow to
distinguish between these two alternatives.

\noindent
{\bf{ACKNOWLEDGMENT}}

The author would like to thank Carlton Caves, David Deutsch, Bill Unruh,
John Archibald Wheeler and Bill Wootters for many useful and encouraging
discussions on the subject of the paper.  This work was supported by NSF
grant No. PHY78-26592, Center for Theoretical Physics of the University of
Texas at Austin, SERC grant to the Department of Astrophysics at Oxford
University and by the Tolman Fellowship at Caltech.

\noindent
{\bf{REFERENCES}}

\noindent
[Note:  References were not updated.]

\begin{enumerate}
\item N. Bohr, The quantum postulate and the recent development of atomic
theory, \underline{Nature} 121:580-590 (1928).
\item W. Heisenberg, Uber den anschaulichen Inhalt der quantentheoretischen
Kinematik und Mechanik, \underline{Z. Physik} 43:172-198 (1927); English
translation, The physical content of quantum kinematic and mechanics, to
appear in: ``Quantum Theory and Measurement,'' J.A. Wheeler and W. H.
Zurek, eds., Princeton University Press (1983).
\item J. von Neumann, ``Mathematical Foundations of Quantum Mechanics,''
translated by R. T. Beyer, Princeton University Press (1955).
\item A. Einstein, B. Podolsky and N. Rosen, Can quantum-mechanical
description of physical reality be considered complete?, Phys. Rev.
47:777-780 (1935).
\item E. Schr\"{o}dinger, Die gegenw\"{a}rtige Situation in der
Quantenmechanik, \underline{Naturwiss}. 23:807-812; 823-828; 844-849
(1935); English translation, The present situation in quantum mechanics, by
J. D. Trimmer in:  \underline{Proc. Am. Phil. Soc.} 124:323-338 (1980).
\item N.  Bohr, Discussion with Einstein on epistemological problems in
atomic physics, \underline{in}: ``Albert Einstein: Philosopher Scientist,''
P. A. Schilpp, ed., The Library of Living Philosophers, Evanston (1949).
\item F. London and E. Bauer, ``La th\'{e}orie de l'observation en
m\'{e}canique quantique,'' Hermann, Paris (1939); English translation ``The
Theory of Observation in Quantum Mechanics'' to appear in ``Quantum Theory
and Measurement,'' J. A. Wheeler and W. H. Zurek, eds., Princeton University
Press (1983).
\item H. D. Zeh, On the irreversibility of time and observation in quantum
theory, in:  ``Foundations of Quantum Mechanics,'' B. D'Espagnat, ed.,
Academic Press, New York (1971).
\item E.  P. Wigner, Review of the quantum mechanical measurement problem,
preprint based on a lecture delivered by E. P. Wigner at Los Alamos, June
1981.  See also E. P. Wigner, this volume, and references therein.
\item W. H. Zurek, Pointer basis of quantum apparatus:  Into what mixture
does the wave packet collapse?, \underline{Phys. Rev. D} 24:1516-1525 (1981).
\item D. Bohm, ``Quantum Theory'', Prentice-Hall, New York (1951).
\item V. B. Braginsky and A. B. Manukin, ``Measurement of Weak Forces in
Physics Experiments,'' University of Chicago Press, Chicago (1977).
\item K. S. Thorne, R. W. P. Drever, C. M. Caves, M. Zimmermann, and V. D.
Sandberg,. Quantum nondemoltion measurements of harmonic oscillators,
\underline{Phys. Rev. Lett.} 40:667-671 (1978).
\item C. M. Caves, K. S. Thorne, R. W. P. Drever, V. D. Sandberg, and M.
Zimmermann, On the measurement of a weak classical force coupled to a
quantummechanical oscillator, \underline{Rev. Mod. Phys.} 52:341-392
(1980); see also C. M. Caves, this volume, and references therein.
\item W. G. Unruh, Quantum nondemolition and gravity wave detection,
\underline{Phys. Rev. D} 19:2888-2896 (1979), see also W. G. Unruh, this
volume.
\item H. Everett III, ``Relative State'' formulation of quantum mechanics,
\underline{Rev. Mod. Phys.} 29:454-462 (1957).
\item H. P. Yuen, this volume.
\item I. Prigogine, C. George, F. Henin, and L. Rosenfeld, A unified
formulation of dynamics and thermodynamics, \underline{Chemica Scripta}
4:5-32 (1973).
\item S. W. Hawking, Breakdown of predictability in gravitational collapse,
\underline{Phys. Rev. D} 14:2460-2473 (1976).
\item R. Penrose, Time asymmetry and quantum gravity, to appear in
``Quantum Gravity II, A Second Oxford Symposium,'' Oxford University Press,
Oxford (1982).
\item E. P. Wigner, Remarks on the Mind~Body question, \underline{in} ``The
Scientist Speculates,'' I. J. Good, ed., Heinmann, London (1961).
\item M. O. Scully, R. Shea and J. D. McCullen, State reduction in quantum
mechanics:  A calculational example, \underline{Phys. Reports} 43:485-498
(1978).
\item M. O. Scully, this volume, and references therein.
\item J. A. Wheeler, The ``past'' and the ``delayed choice'' double slit
experiment, \underline{in}:  ``Mathematical Foundations of Quantum
Theory,'' Academic Press, New York (1978).
\item J. A. Wheeler, Frontiers of time, in:  ``Problems in the Foundations
of Physics'', N. Toraldo di Francia, ed., North Holland, Amsterdam (1979).
\item W. K. Wootters and W. H. Zurek, Complementarity in the double-slit
experiment:  Quantum nonseparability and a quantitative statement of Bohr's
principle, \underline{Phys. Rev. D} 19:473-484 (1979).
\item W. C. Wickes, C. 0. Alley, and 0. Jakubowicz, A ``delayed choice''
quantum mechanics experiment, University of Maryland preprint (1981); see
also L. S. Bartell, Complementarity in the double-slit experiment:  On
simple realizable systems for observing intermediate particle-wave
behavior, \underline{Phys. Rev. D} 21:1698-1699 (1980).
\item H. S. Green, Observation in quantum mechanics, \underline{Nuovo
Cimento} 9:880-889 (1958).
\item A. Daneri, A. Loinger and G. M. Prosperi, Quantum theory of
measurement and ergodicity conditions, \underline{Nuclear Phys.} 33:297-319
(1962).
\item K. Hepp, Quantum theory of measurement and macroscopic observables,
\underline{Helv. Phys. Acta} 45:237-248 (1972).
\item F. Haake and W. Weidlich, A model for measuring process in quantum
theory, Z.Phys. 213:451-465 (1968).
\item G. G. Emch, On quantum measurement processes, \underline{Helv. Phys.
Acta} 45:10149-1056 (1972).
\item J. S. Bell, On wave packet reduction in the Coleman-Hepp model,
\underline{Helv. Phys. Acta} 48:93-98 (1975).
\item R. W. Zwanzig, Statistical mechanics of irreversibility,
\underline{in} ``Lectures on Theoretical Physics'' 3:106-141,
Wiley-Interscience, New York (1961).
\item H. D. Zeh, On the interpretation of measurement in quantum theory,
\underline{Found Phys.} 1:69-76 (1970).
\item B. S. DeWitt and N. Graham, eds., "The Many-World Interpretation of
Quantum Mechanics" Princeton University Press, Princeton (1973).
\item C. H. Shannon and W. Weaver, ``The Mathematical Theory of
Communication,'' University of Illinois Press, Urbana (1949).
\item B. D'Espagnat, ``Conceptual Foundations of Quantum Mechanics''
Benjamin, Reading, Mass. (second edition, 1976).
\item R. Balescu, Chapters 14-17 in ``Equilibrium and Nonequilibrium
Statistical Mechanics,'' Wiley-Interscience, New York (1975).
\item D. Deutsch, Quantum theory as a universal physical theory,
\underline{Phys. Rep.}, submitted.
\item P. Bocchieri and A. Loinger, Quantum recurrence theorem, 
\underline{Phys. Rev.}
107:337-338 (1957).
\item L. Lugiato, this volume.
\item S. Stenholm, this volume.
\item S. Machida and M. Namiki, Theory of measurement in quantum mechanics;
\underline{Progr. Theor. Phys.} 63:1457-1473 and 1833-1847 (1980).
\item H. Araki, A remark on Machida-Namiki theory of measurement,
\underline{Progr.}
\underline{Theor. Phys.} 64:719-730 (1980).
\item N. Bohr, Can quantum-mechanical description of physical reality be
considered complete?, \underline{Phys. Rev.} 48:696-702 (1935).
\end{enumerate}
\end{document}